# Teleportation of two-particle entangled state via cluster state

Da-Chuang Li and Zhuo-Liang Cao

*Key Laboratory of Opto-electronic Information Acquisition and Manipulation*

*Ministry of Education, and School of Physics and Material Science, Anhui University,*

*Hefei 230039, PR China*

**Abstract.** We present two schemes for teleporting an unknown two-particle entangled state from a sender (Alice) to a receiver (Bob) via a four-particle entangled cluster state. As it is shown, the unknown two-particle entangled state can be teleported perfectly, and the successful possibilities and fidelities of the two schemes both reach unit.

1. Introduction

Quantum entanglement plays a more and more important role in quantum information science because its nonlocality properties. Entanglement is considered as the fundamental resource of quantum information processing such as quantum teleportation [1-6], quantum dense coding [7-9], quantum secret sharing [10-12], and so on. Quantum teleportation, first proposed by Bennett *et al*. [1] and experimentally realized by Bouwmeester *et al*. [2] and Boschi *et al*. [3], can transmit an unknown quantum state from a sender to a receiver at a distant location via a quantum channel with the help of some classical information. Furthermore distributed entangled states make it possible to send an unknown state to a long distance.

The entangled state of three qubits can be classified into GHZ and W class state [13]. A number of applications using these two kinds of state in quantum information have been proposed, for example, Shi *et al*. [14] have proposed a scheme to teleport a two-particle state with a GHZ class state, and we have proposed a scheme [15] to teleport the two-particle state with the W class state. In the above two proposals, both can be successfully realized with a certain probability if the receiver performs an appropriate unitary operation, but both of the schemes need to introduce an auxiliary qubit, furthermore, what they have teleported is not an arbitrary bipartite entangled state.

Entanglement in four qubits is more complicated than that in three qubits. In Ref. [16], they show that cluster states have some particular characters in the case of $N > 3$, for instance, the cluster states have the properties both of the GHZ-class and the W-class entangled states, and they are harder to be destroyed by local operations than GHZ-class states, etc. We present two teleportation schemes, where the teleportation of two-particle entangled state, both in special and arbitrary cases, can be realized via a four-particle cluster state. One is the teleportation of a special bipartite entangled state; the other is the teleportation of an arbitrary two-particle entangled state. The successful possibilities and fidelities of our schemes are both reach 1.0.

## 2. Teleportation of a special bipartite entangled state

In this scheme, suppose the state to be teleported has the following form

$$|\Psi\rangle_{12} = \alpha|00\rangle_{12} + \delta|11\rangle_{12}, \tag{1}$$

where $\alpha$ and $\delta$ are unknown parameters, and $|\alpha|^2 + |\delta|^2 = 1$. A cluster state is used as quantum channel between Alice and Bob, which is in the following state

$$|\Psi\rangle_{3456} = \frac{1}{2}\left(|0000\rangle_{3456} + |0011\rangle_{3456} + |1100\rangle_{3456} - |1111\rangle_{3456}\right), \tag{2}$$

particles 1, 2, 3 and 6 belong to Alice, particles 4 and 5 belong to Bob. All the subscripts denote the ordinal number of particles. Thus the total state of the system can be expressed as

$$|\Psi\rangle = \frac{1}{2}\left(\alpha|00\rangle_{12} + \delta|11\rangle_{12}\right) \otimes \left(|0000\rangle_{3456} + |0011\rangle_{3456}\right.$$
$$\left. + |1100\rangle_{3456} - |1111\rangle_{3456}\right). \tag{3}$$

In order to realize the teleportation, Alice first performs BSMs on particles (1, 3) and particles (2, 6), respectively. Then the state of particles 4 and 5 collapses into one of the following states

$$_{26}\langle\Phi^\pm|_{13}\langle\Phi^\pm|\Psi\rangle = \frac{1}{4}\left(\alpha|00\rangle_{45} \pm^1 \mp^2 \delta|11\rangle_{45}\right), \tag{4a}$$

$$_{26}\langle\Psi^\pm|_{13}\langle\Phi^\pm|\Psi\rangle = \frac{1}{4}\left(\alpha|01\rangle_{45} \pm^1 \pm^2 \delta|10\rangle_{45}\right), \tag{4b}$$

$$_{26}\langle\Phi^\pm|_{13}\langle\Psi^\pm|\Psi\rangle = \frac{1}{4}\left(\alpha|10\rangle_{45} \pm^1 \pm^2 \delta|01\rangle_{45}\right), \tag{4c}$$

$$_{26}\langle\Psi^{\pm}|_{13}\langle\Psi^{\pm}|\Psi\rangle = \frac{1}{4}\left(-\alpha|11\rangle_{45} \pm^1 \pm^2 \delta|00\rangle_{45}\right), \quad (4d)$$

where $|\Phi^{\pm}\rangle_{ij} = 1/\sqrt{2}(|0\rangle_i|0\rangle_j \pm |1\rangle_i|1\rangle_j)$ and $|\Psi^{\pm}\rangle_{ij} = 1/\sqrt{2}(|0\rangle_i|1\rangle_j \pm |1\rangle_i|0\rangle_j)$ are four Bell states, $\pm^1(\mp^1)$ and $\pm^2(\mp^2)$ denote the results corresponding to BSMs on particle pairs (1, 3) and (2, 6), respectively. After that, Alice tells Bob her measurement results via a classical channel. Finally, Bob can obtain the unknown state on particles 4 and 5 by performing appropriate unitary transformations. We discuss the operations in detail below.

Without loss of generality, if Alice's measurement results are $|\Phi^+\rangle_{13}$ and $|\Phi^+\rangle_{26}$ respectively, then the particle pair (4, 5) is collapsed into the state $\alpha|00\rangle_{45} - \delta|11\rangle_{45}$. Bob need perform $U_{45} = I_4 \otimes Z_5$ or $U_{45} = Z_4 \otimes I_5$ on the particle pair (4, 5) to reconstruct the original state, where $I$ and $Z$ are the identity operator and the Pauli operator $\sigma_z$, respectively. For other measurement results, similarly, Bob should perform appropriate operations on particles 4 and 5, which are shown in table 1. By calculating, the total successful probability of the teleportation is $P_{total} = \left(\frac{1}{4}\right)^2 \times 16 = 1$.

Table 1: The unitary transformations corresponding to Alice's measurement results.

| Alice's result | Bob's operations | Alice's result | Bob's operations |
| --- | --- | --- | --- |
| $|\Phi^+\rangle_{13}|\Phi^+\rangle_{26}$ | $I_4 \otimes Z_5$ or $Z_4 \otimes I_5$ | $|\Psi^+\rangle_{13}|\Phi^+\rangle_{26}$ | $X_4 \otimes I_5$ |
| $|\Phi^+\rangle_{13}|\Phi^-\rangle_{26}$ | $I_4 \otimes I_5$ | $|\Psi^+\rangle_{13}|\Phi^-\rangle_{26}$ | $X_4 \otimes Z_5$ or $Y_4 \otimes I_5$ |
| $|\Phi^+\rangle_{13}|\Psi^+\rangle_{26}$ | $I_4 \otimes X_5$ | $|\Psi^+\rangle_{13}|\Psi^+\rangle_{26}$ | $X_4 \otimes Y_5$ or $Y_4 \otimes X_5$ |
| $|\Phi^+\rangle_{13}|\Psi^-\rangle_{26}$ | $I_4 \otimes Y_5$ or $Z_4 \otimes X_5$ | $|\Psi^+\rangle_{13}|\Psi^-\rangle_{26}$ | $X_4 \otimes X_5$ |
| $|\Phi^-\rangle_{13}|\Phi^+\rangle_{26}$ | $I_4 \otimes I_5$ | $|\Psi^-\rangle_{13}|\Phi^+\rangle_{26}$ | $X_4 \otimes Z_5$ or $Y_4 \otimes I_5$ |
| $|\Phi^-\rangle_{13}|\Phi^-\rangle_{26}$ | $I_4 \otimes Z_5$ or $Z_4 \otimes I_5$ | $|\Psi^-\rangle_{13}|\Phi^-\rangle_{26}$ | $X_4 \otimes I_5$ |
| $|\Phi^-\rangle_{13}|\Psi^+\rangle_{26}$ | $I_4 \otimes Y_5$ or $Z_4 \otimes X_5$ | $|\Psi^-\rangle_{13}|\Psi^+\rangle_{26}$ | $X_4 \otimes X_5$ |
| $|\Phi^-\rangle_{13}|\Psi^-\rangle_{26}$ | $I_4 \otimes X_5$ | $|\Psi^-\rangle_{13}|\Psi^-\rangle_{26}$ | $X_4 \otimes Y_5$ or $Y_4 \otimes X_5$ |

## 3. Teleportation of an arbitrary two-particle entangled state

In this section, we will teleport an arbitrary two-particle entangled state. We assume the state to be teleported is

$$|\Psi\rangle_{12} = \alpha|00\rangle_{12} + \beta|01\rangle_{12} + \gamma|10\rangle_{12} + \delta|11\rangle_{12}, \quad (5)$$

where $\alpha, \beta, \gamma$ and $\delta$ are unknown parameters, and $|\alpha|^2 + |\beta|^2 + |\gamma|^2 + |\delta|^2 = 1$. The quantum channel is also the cluster state in Eq. (2). Particles 1, 2, 3 and 6 belong to Alice, and particles 4 and 5 belong to Bob. Thus the total state of the system can be expressed as

$$|\Psi\rangle = \frac{1}{2}\left(\alpha|00\rangle_{12} + \beta|01\rangle_{12} + \gamma|10\rangle_{12} + \delta|11\rangle_{12}\right)$$
$$\otimes \left(|0000\rangle_{3456} + |0011\rangle_{3456} + |1100\rangle_{3456} - |1111\rangle_{3456}\right). \quad (6)$$

To realize the teleportation, BSMs on particles (1, 3) and particles (2, 6) are made by Alice at the first step, and then all the 16 possible collapsed states of particles (4, 5) are

$$_{26}\langle\Phi^\pm|_{13}\langle\Phi^\pm|\Psi\rangle = \frac{1}{4}\left(\alpha|00\rangle_{45} \pm^2 \beta|01\rangle_{45} \pm^1 \gamma|10\rangle_{45} \pm^1 \mp^2 \delta|11\rangle_{45}\right), \quad (7a)$$

$$_{26}\langle\Psi^\pm|_{13}\langle\Phi^\pm|\Psi\rangle = \frac{1}{4}\left(\alpha|01\rangle_{45} \pm^2 \beta|00\rangle_{45} \mp^1 \gamma|11\rangle_{45} \pm^1 \pm^2 \delta|10\rangle_{45}\right), \quad (7b)$$

$$_{26}\langle\Phi^\pm|_{13}\langle\Psi^\pm|\Psi\rangle = \frac{1}{4}\left(\alpha|10\rangle_{45} \mp^2 \beta|11\rangle_{45} \pm^1 \gamma|00\rangle_{45} \pm^1 \pm^2 \delta|01\rangle_{45}\right), \quad (7c)$$

$$_{26}\langle\Psi^\pm|_{13}\langle\Psi^\pm|\Psi\rangle = \frac{1}{4}\left(-\alpha|11\rangle_{45} \pm^2 \beta|10\rangle_{45} \pm^1 \gamma|01\rangle_{45} \pm^1 \pm^2 \delta|00\rangle_{45}\right). \quad (7d)$$

Similarly, where $|\Phi^\pm\rangle_{ij}$ and $|\Psi^\pm\rangle_{ij}$ are the Bell states of the particle pair (i, j), $\pm^1 (\mp^1)$ and $\pm^2 (\mp^2)$ denote the results corresponding to BSMs on particle pairs (1, 3) and (2, 6), respectively.

Then Bob performs a quantum controlled phase gate operation on the particles 4 and 5, where the particle 4 is the control bit and the particle 5 is the target bit, *i.e.*, if and only if particle 4 is in the state $|1\rangle$, particle 5 is performed an operation of Pauli operator ($\sigma_z$). Thus the Eq. (8) becomes

$$_{26}\langle\Phi^\pm|_{13}\langle\Phi^\pm|\Psi\rangle = \frac{1}{4}\left(\alpha|00\rangle_{45} \pm^2 \beta|01\rangle_{45} \pm^1 \gamma|10\rangle_{45} \pm^1 \pm^2 \delta|11\rangle_{45}\right), \quad (8a)$$

$$_{26}\langle\Psi^\pm|_{13}\langle\Phi^\pm|\Psi\rangle = \frac{1}{4}\left(\alpha|01\rangle_{45} \pm^2 \beta|00\rangle_{45} \pm^1 \gamma|11\rangle_{45} \pm^1 \pm^2 \delta|10\rangle_{45}\right), \quad (8b)$$

$$_{26}\langle\Phi^\pm|_{13}\langle\Psi^\pm|\Psi\rangle = \frac{1}{4}\left(\alpha|10\rangle_{45} \pm^2 \beta|11\rangle_{45} \pm^1 \gamma|00\rangle_{45} \pm^1 \pm^2 \delta|01\rangle_{45}\right), \quad (8c)$$

$$_{26}\langle\Psi^{\pm}|_{13}\langle\Psi^{\pm}|\Psi\rangle = \frac{1}{4}\left(\alpha|11\rangle_{45} \pm^{2} \beta|10\rangle_{45} \pm^{1} \gamma|01\rangle_{45} \pm^{1} \pm^{2} \delta|00\rangle_{45}\right). \tag{8d}$$

After knowing the four bits classical information from Alice, Bob will perform relevant unitary transformation to reproduce the unknown state on particles 4 and 5. Without loss of generality, if the measurement results of particle pairs (1, 3) and (2, 6) are $|\Phi^{+}\rangle_{13}$ and $|\Phi^{-}\rangle_{26}$, respectively, particles 4 and 5 will collapse into the state $\alpha|00\rangle_{45} - \beta|01\rangle_{45} + \gamma|10\rangle_{45} - \delta|11\rangle_{45}$. Then Bob performs $U_{45} = I_4 \otimes Z_5$ on particles 4 and 5 to reconstruct the original state. The others possible cases are described in table 2. Thus the arbitrary two-particle entangled state can be reproduced on Bob's side successfully. The total probability of success is $P_{total} = \left(\frac{1}{4}\right)^2 \times 16 = 1$, and the fidelity of the output state is 1.0.

Table 2: The unitary transformations corresponding to Alice's measurement results.

| Alice's result | Bob's operations | Alice's result | Bob's operations |
| --- | --- | --- | --- |
| $\|\Phi^{+}\rangle_{13}\|\Phi^{+}\rangle_{26}$ | $I_4 \otimes I_5$ | $\|\Psi^{+}\rangle_{13}\|\Phi^{+}\rangle_{26}$ | $X_4 \otimes I_5$ |
| $\|\Phi^{+}\rangle_{13}\|\Phi^{-}\rangle_{26}$ | $I_4 \otimes Z_5$ | $\|\Psi^{+}\rangle_{13}\|\Phi^{-}\rangle_{26}$ | $X_4 \otimes Z_5$ |
| $\|\Phi^{+}\rangle_{13}\|\Psi^{+}\rangle_{26}$ | $I_4 \otimes X_5$ | $\|\Psi^{+}\rangle_{13}\|\Psi^{+}\rangle_{26}$ | $X_4 \otimes X_5$ |
| $\|\Phi^{+}\rangle_{13}\|\Psi^{-}\rangle_{26}$ | $I_4 \otimes Y_5$ | $\|\Psi^{+}\rangle_{13}\|\Psi^{-}\rangle_{26}$ | $X_4 \otimes Y_5$ |
| $\|\Phi^{-}\rangle_{13}\|\Phi^{+}\rangle_{26}$ | $Z_4 \otimes I_5$ | $\|\Psi^{-}\rangle_{13}\|\Phi^{+}\rangle_{26}$ | $Y_4 \otimes I_5$ |
| $\|\Phi^{-}\rangle_{13}\|\Phi^{-}\rangle_{26}$ | $Z_4 \otimes Z_5$ | $\|\Psi^{-}\rangle_{13}\|\Phi^{-}\rangle_{26}$ | $Y_4 \otimes Z_5$ |
| $\|\Phi^{-}\rangle_{13}\|\Psi^{+}\rangle_{26}$ | $Z_4 \otimes X_5$ | $\|\Psi^{-}\rangle_{13}\|\Psi^{+}\rangle_{26}$ | $Y_4 \otimes X_5$ |
| $\|\Phi^{-}\rangle_{13}\|\Psi^{-}\rangle_{26}$ | $Z_4 \otimes Y_5$ | $\|\Psi^{-}\rangle_{13}\|\Psi^{-}\rangle_{26}$ | $Y_4 \otimes Y_5$ |

**4. Conclusion**

In this paper, two different schemes for teleporting an unknown two-particle entangled state are proposed. Here we use a four-particle cluster state as the quantum channel. In the first scheme, we teleport a bipartite entangled state, the receiver Bob can reconstruct the teleported state according to Alice's measurement results, and the successful possibility and fidelity are both 1.0. In the latter scheme we teleport an arbitrary and unknown two-particle entangled state. Contrast to the first scheme, in order to realize the teleportation, Bob should perform a quantum controlled phase gate operation besides the unitary transformation. The advantage of this scheme is that the state teleported is a more general state. Also, the successful possibility and fidelity of this scheme both reach 1.0.

**Acknowledgements**

This work is supported by the Key Program of the Education Department of Anhui Province under Grant No: 2006KJ070A, the Program of the Education Department of Anhui Province under Grant No: 2006KJ057B and the Talent Foundation of Anhui University.